\documentclass[twocolumn,showpacs,prc]{revtex4}       

\newcommand{ \be }{\begin{equation}}    
\newcommand{ \ee }{\end{equation}}    
\newcommand{ \bea }{\begin{eqnarray}}    
\newcommand{ \eea }{\end{eqnarray}}

\usepackage{graphicx}
\usepackage{dcolumn}
\usepackage{bm}
\usepackage{verbatim}
\usepackage{amsmath}
\usepackage[usenames, dvipsnames]{color}

\begin{document}       
\voffset=0.5 in    
       
       
\title{    
\begin{flushright}    
\end{flushright}    
Procedure for measuring photon and vector meson circular polarization variation with respect to the reaction plane in relativistic heavy-ion collisions
}    
       
\author{A.H. Tang$^1$, G.Wang$^2$}
\affiliation{$^1$Brookhaven National Laboratory, Upton, New York 11973 \\
$^2$Department of Physics and Astronomy, University of California, Los Angeles, California 90095 \\
}
       
    
\begin{abstract}       
The electromagnetic (EM) field pattern created by spectators in relativistic heavy-ion collisions plants a seed of
positive (negative) magnetic helicity in the hemisphere above (below) the reaction plane.
Owing to the chiral anomaly, the magnetic helicity interacts with the fermionic helicity of the collision system,
and causes photons emitted in upper- and lower-hemispheres to have different preferences in the circular polarization. 
Similar helicity separation for massive particles, due to the global vorticity, is also possible. In this paper, we lay down a procedure to measure the variation of the circular polarization w.r.t the reaction plane 
in relativistic heavy-ion collisions for massless photons, as well as similar polarization patterns for vector mesons decaying into two daughters. We propose to study the yield differentially and compare the yield 
between upper- and lower-hemispheres in order to identify and quantify such effects.
\end{abstract}       
       
\pacs{25.75.Ld}       
    
\maketitle       

\section{Introduction}

In a non-central relativistic heavy-ion collision, an ultra-strong magnetic field will be produced inside the
participant zone by energetic spectator protons~\cite{KharzeevMcLerranWarringa}.
On average, the magnetic field is perpendicular to the so-called ``reaction plane" (RP) that contains 
the impact parameter and the beam momenta.  
This strong magnetic field, coupled with quantum anomalies, will cause a series of  
novel non-dissipative transport effects that could survive the expansion of the fireball and be detected in experiments. 
For a review on these subjects, see Ref.~\cite{CMEReview2015}. 
Furthermore, the initial magnetic helicity of the collision system can be quite large,
and bears opposite signs in the upper- and lower-hemispheres.
It was recently realized~\cite{HironoKharzeevYin,Tuchin2015,Manuel2015,McLerran2014} that owing to the chiral anomaly,
the helicity can be transferred back and forth between the magnetic flux and fermions as the collision system evolves,
so that the magnetic helicity could last long enough to yield photons with opposite circular polarizations 
in the hemispheres above and below the RP~\cite{HironoKharzeevYin,Ipp2008,Mamo2013}. A similar asymmetry in photon polarization can also result from the initial global quark polarization~\cite{Wang2005} which could effectively lead to a polarization of photons~\cite{Ipp2008}.
This local imbalance of photon circular polarization could be observed in experiments, e.g., by studying the polarization preference w.r.t the RP for photons that
convert into $e^+e^-$ pairs. In this paper, we discuss a few practical thoughts of carrying out such a measurement, 
first for photons and then for vector mesons.

\section{Photons}
Photon circular polarization can be measured with dedicated polarimeters~\cite{Trautmann1980}, 
by studying the count rate asymmetries with a local magnet. Such a setup is infeasible for large-scale heavy-ion experiments 
like the Relativistic Heavy Ion Collider or the Large Hadron Collider, as the local magnet will complicate the magnetic field 
that is used for the track detection of charged particles. Nevertheless, one can still probe photon circular polarization 
by tracking $e^+e^-$ pairs from photon conversion.

\begin{figure}[h]       
\centering
\makebox[2cm]{\includegraphics[width=0.45 \textwidth]{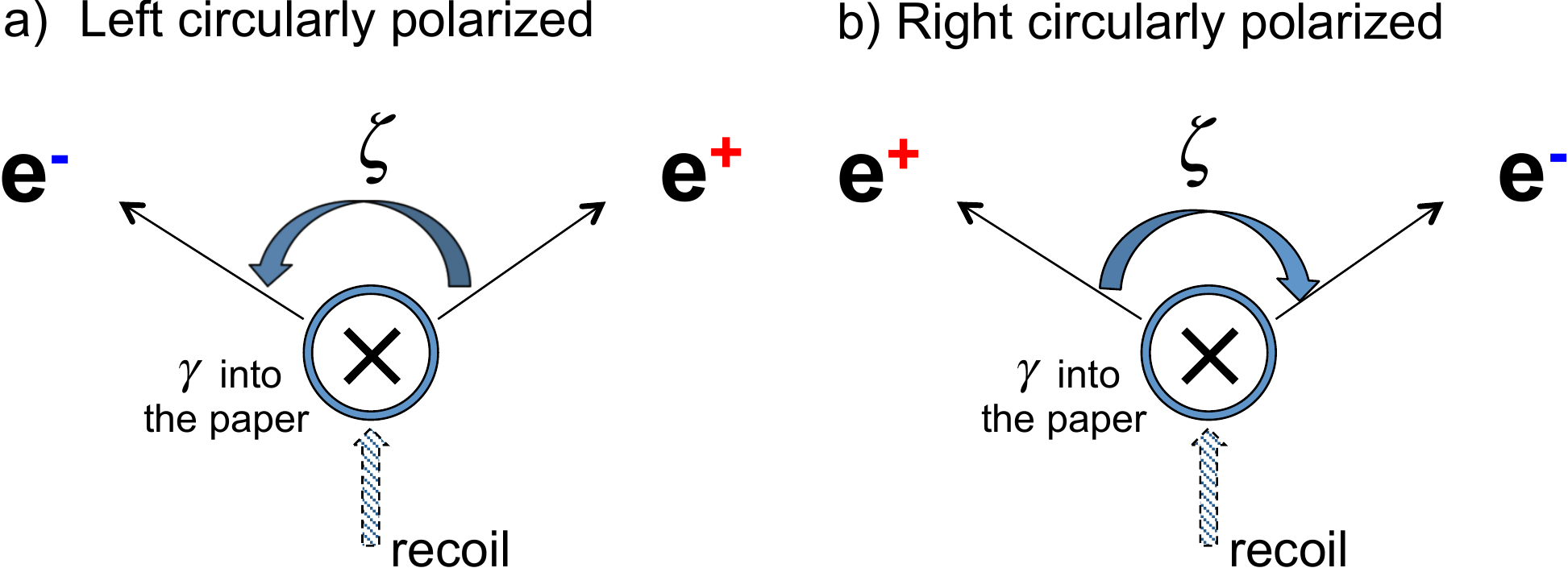}}
\caption{A left (a) or right (b) circularly polarized photon is converted into an electron-positron pair 
with the help of the recoil of an atomic nucleus.}
  \label{fig:Cartoon1}
\end{figure}

A photon will not decay by itself. However, when a photon is near an atomic nucleus, its energy can be converted 
into an electron-position pair. Panels a) and b) in Fig.~\ref{fig:Cartoon1} show the typical conversion process for left and 
right circularly polarized photons, respectively. Here $\zeta$ is the angle of the electron relative to the positron, when both 
are projected onto the plane perpendicular to the motion direction of the photon. Owing to the opposite circular polarizations, 
the preferred direction orderings between the emitted electron and positron are opposite in the two cases.

The cross-section for the pair production by completely circularly polarized high-energy photons 
is given as~\cite{OlsenMaximon59,OlsenMaximon62}
\bea
d\sigma \propto 1 \pm a \cdot \mathrm{sin} \zeta, 
\eea
where the $+$ and $-$ signs are for right and left circularly polarized photons, respectively. 
$a$ ($\in [0,1]$) is a positive quantity dependent on the photon energy and the energies and momenta of its two daughters, 
as well as the atomic number $Z$ of the recoiling nucleus.

In the presence of the magnetic helicity effect, the yields of left and right circularly polarized photons can each be
parametrized as a sine wave modulation w.r.t. the RP, which is similar to the directed flow ($v_1$)~\cite{PoskanzerVoloshin} 
that is modulated by a cosine wave.
Furthermore, the sine coefficients have the same magnitude and opposite signs for left and right circularly polarized photons. 
With this parametrization scheme, the effect of the magnetic helicity aforementioned is strongest in the direction 
perpendicular to the RP (along the direction of the magnetic field), and weakest in the RP, and so is the difference 
in yield between the two groups of photons. 
The cross-sections for right and left circularly polarized photons can be written separately as
\bea
d\sigma_R &\propto& ( 1 + a \cdot \mathrm{sin} \zeta )( 1 + b \cdot \mathrm{sin} \phi ) \\
d\sigma_L &\propto& ( 1 - a \cdot \mathrm{sin} \zeta )( 1 - b \cdot \mathrm{sin} \phi ),
\eea
where $\phi$ is the photon's azimuthal angle relative to the RP, and $b$ ($\in [0,1]$) represents the magnitude of the modulation. 
The total cross-section is the sum of $d\sigma_R$ and $d\sigma_L$:
\bea
d\sigma \propto 1 + H \mathrm{sin} \zeta \mathrm{sin} \phi,
\eea
where $H \equiv ab$, and $H \in [0,1]$.

$\zeta$, $\phi$ are both experimentally measurable quantities, so in principle 
$H$ can be obtained from the cross-section measurement. 
A finite $H$ implies a finite $b$ or a finite effect of modulation w.r.t the RP due to the external magnetic field. 
Note that only the modulation w.r.t RP carries the information of the asymmetry between left and right circularly polarized photons,
so other sources of photons (mostly from hadron decays) do not contribute to $H$ as long as they are produced symmetrically 
w.r.t RP in terms of the left-right circular polarization. Generally speaking, photons can have other finite Fourier coefficients, e.g., $v_1$ and $v_2$~\cite{PoskanzerVoloshin}, but they do not couple with $H$ as long as we assume those coefficients are the same 
for left and right circularly polarized photons.

The (in)efficiency as a function of $\phi$ angle can be easily compensated in experiments, 
as long as the RP distribution and the azimuthal angle distribution of photons in the laboratory frame are flattened, 
which is a common procedure when analyzing anisotropic flow~\cite{PoskanzerVoloshin}.  
We also assume that the efficiency as a function of $\zeta$ can be measured by usual 
Monte Carlo studies with individual experiment setups. Taking the difference in yield-probability between the upper hemisphere 
and the lower hemisphere, 
\bea
\nonumber \Delta Y_{\rm{up}-\rm{down}}^\gamma & = & \int_0^\pi  \mathrm{d}\sigma \mathrm{d}\phi - \int_\pi^{2\pi} \mathrm{d}\sigma \mathrm{d}\phi\\ 
& \propto & 4H\mathrm{sin}\zeta
\eea
and the difference between the left hemisphere and right hemisphere (separated by $\phi=\pi/2$) would be zero 
owing to the symmetry of the system: 
\bea
\nonumber \Delta Y_{\rm{left}-\rm{right}}^\gamma & = & \int_{\frac{\pi}{2}}^{\frac{3\pi}{2}}  \mathrm{d}\sigma \mathrm{d}\phi - \int_{-\frac{\pi}{2}}^{\frac{\pi}{2}} \mathrm{d}\sigma \mathrm{d}\phi\\ 
& = & 0,
\eea
which can be regarded as a consistency check in experiments. Note that when there is no signal, 
$\Delta Y_{\rm{up}-\rm{down}}^\gamma = \Delta Y_{\rm{left}-\rm{right}}^\gamma =0$, 
and a finite efficiency $\epsilon(\zeta)$ will not cause an artificial, finite $\Delta Y_{\rm{up}-\rm{down}}^\gamma$. 

In heavy-ion collider experiments, photons are usually reconstructed at secondary vertices 
where they strike detector materials and emit $e^+e^-$ pairs.
The photon sample is inevitably contaminated by combinatorial background formed by random $e^+e^-$ pairs. 
However, the background contribution should be symmetric w.r.t both $\phi=0$ and $\phi=\pi/2$ plane, and 
be canceled in this procedure by taking the difference in yield between up and down (or left and right) hemispheres. 
The same argument also holds for the study of vector mesons discussed below.

\section{ vector mesons }

Unlike photons that can be only transversely polarized (with the spin direction either parallel or anti-parallel 
to the momentum direction), vector mesons can be both longitudinally and transversely polarized. 
Below, we describe a procedure to measure a potential difference, if a similar effect exists, between left and right 
circular polarizations for vector mesons, as well as a way to study the variation of such a difference w.r.t the RP. The procedure is 
applicable to vector mesons decaying into two daughters, such as $\omega$, $\phi$, $J/\psi$ etc, and applicable to virtual photons decaying into two leptons by firstly fluctuating into vector mesons.  The earlier the particle is produced, the better it experiences the magnetic helicity that happens at the very early time (a few $fm/c$) of the collision. In addition, the lighter the particle, the easier for it to be polarized by the magnetic field. In the end what particle to choose for this study is a balanced decision between the particle's production time, mass, the purity of the data sample, and the statistics available. The latter two factors vary from one experiment to another. Note that although here we mentioned the magnetic helicity, the   system can have helicity separation originated from vorticity~\cite{Wang2005,Baznat2013,Deng2016}, which may have an effect on massive particles. 

The polarization of vector mesons can be studied with the distribution of its decay products in azimuthal and polar 
angles (for example, for $J/\psi$ particle see ~\cite{Faccioli2010, Faccioli2011,STARJpsiPolarizationInPP}) in a chosen coordinate system (frame). 
The four popular choices of frame are the particle rest frame~\cite{Faccioli2010}, 
the Gottfried-Jackson frame~\cite{GJ,Faccioli2010}, the center-of-mass helicity frame~\cite{Faccioli2010}, 
and the Collins-Soper frame~\cite{CS}. However, no discussion is given in~\cite{Faccioli2010, Faccioli2011} 
to distinguish the left and right circular polarizations, both belonging to transverse polarization. 

\begin{figure}[h]       
\centering
\makebox[2cm]{\includegraphics[width=0.45 \textwidth]{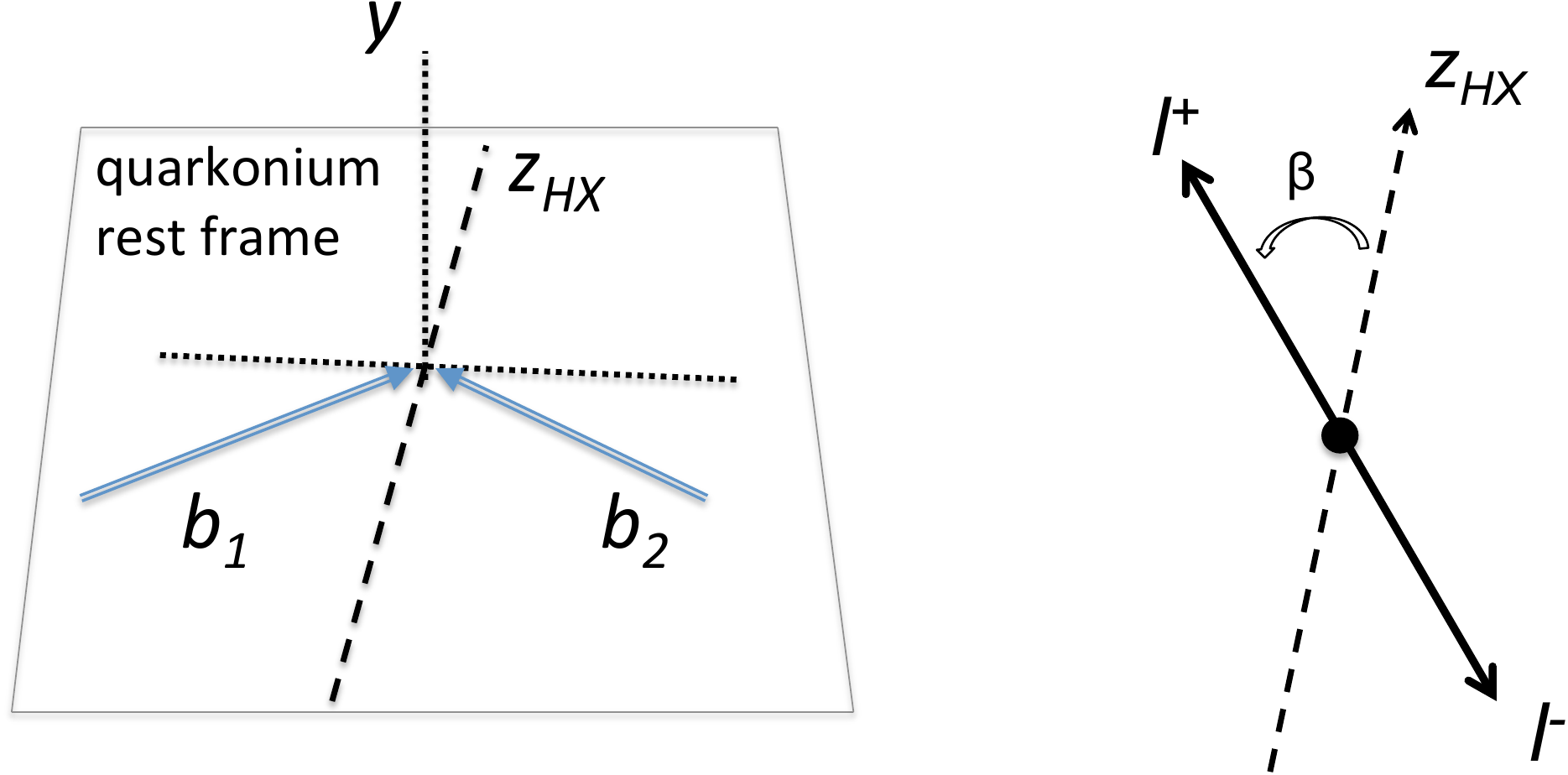}}
\caption{Illustration of the helicity frame (left), in which $b_1$ and $b_2$ are beam directions in the quarkonium rest frame, 
and the orientation of the polar axis ($Z_{HX}$) is set to be the opposite of the direction of motion of the interaction point. 
The polar angle ($\beta$) is defined as the angle between the polar axis and the positive lepton (right).}
  \label{fig:Cartoon2}
\end{figure}

Our discussion will be limited to the center-of-mass helicity frame only. In the helicity frame~\cite{Faccioli2010}, 
the orientation of the polar axis is set to be the opposite of the direction of motion of the interaction point 
(the flight direction of the quarkonium itself in the center-of-mass of the colliding beam). 
The illustration of the helicity frame and the definition of polar angle ($\beta$) is given in Fig.~\ref{fig:Cartoon2}.

As a side remark, the definition of transverse and longitudinal polarization for vector mesons is is commonly used, but counterintuitive. The definition is originated in analogy to a photon which is said to be {\it transversely} polarized with spin projection $J_z=\pm1$, owing to the fact that the EM field carried by a photon oscillates in the transverse plane with respect to its momentum, while its spin is aligned along the momentum.  {\it longitudinal} polarization means $J_z=0$. We respect this convention in this paper.

\subsection{\label{sec:Jpsi1}Asymmetry in yield w.r.t RP for left and right-circular polarization only}

Without loss of generality, let's assume the vector meson under discussion is a $J/\psi$ particle decaying into a $e^+e^-$ pair. 
In the helicity frame, the polar angle distributions of positrons for left-handed, longitudinal and right-handed $J/\psi$ 
can be written as ~\cite{MarkThomson}
\bea
P_{T_l} & \propto & \frac{1}{4}(1-\mathrm{cos}\beta)^2 \\
P_L & \propto & \frac{1}{2} \mathrm{sin^2}\beta \\
P_{T_r} & \propto & \frac{1}{4}(1+\mathrm{cos}\beta)^2.
\eea
If one assumes the left-right symmetry and introduces $\lambda_{\beta} \in [-1,1]$, 
with $\lambda_{\beta} = 1 (-1)$ representing the complete transverse (longitudinal) polarization, 
the polarization-summed cross-section can be expressed as
\bea
P_{pol\_sum} & \propto &  \frac{1 + \lambda_{\beta}}{2} (P_{T_l} + P_{T_r}) + \frac{1-\lambda_{\beta}}{2} P_L  \\
& \propto & 1 + \lambda_{\beta} \mathrm{cos^2}\beta,
\eea
with which we have recovered the formula to describe $J/\psi$ polarization in the polar angle 
in Refs.~\cite{Faccioli2010, Faccioli2011}. 

Now we relax the assumption of the left-right symmetry, and introduce $x = A \mathrm{sin}\phi $, 
where $\phi$ is $J/\psi$'s azimuthal angle relative to the RP, and $A$ ($\in [0,1]$) represents the asymmetry 
in the emission of left and right circularly polarized $J/\psi$ w.r.t RP,
\bea
\nonumber P_{pol\_sum} & \propto & \frac{1 + \lambda_{\beta}}{2} ( \frac{1-x}{2} P_{T_l} + \frac{1+x}{2} P_{T_r}) + \\
  &  & \frac{1-\lambda_{\beta}}{2} P_L. 
\eea 
With this definition, when $x$ is 1 (-1), $J/\psi$'s are produced 100\% with right (left) circular polarization. 
After normalization in $\beta$ space, $P_{pol\_sum}$ is written in full form as :
\bea
\nonumber P_{pol\_sum} & = &  \frac{3}{32}[5+\lambda_\beta+4A(1+\lambda_\beta)\mathrm{sin}\phi\mathrm{cos}\beta \\
+(3\lambda_\beta -1)\mathrm{cos}2\beta]
\eea 

Taking into account the modulation w.r.t. the RP due to the elliptic flow ($v_2$)~\cite{PoskanzerVoloshin}, 
the final probability in yield becomes
\bea
y (\phi,\beta)= \frac{1}{2\pi}(1+2v_2\mathrm{cos}2\phi)P_{pol\_sum}.
 \label{eq:yProb}
\eea 
\begin{figure}[h]       
\centering
\makebox[2cm]{\includegraphics[width=0.3 \textwidth]{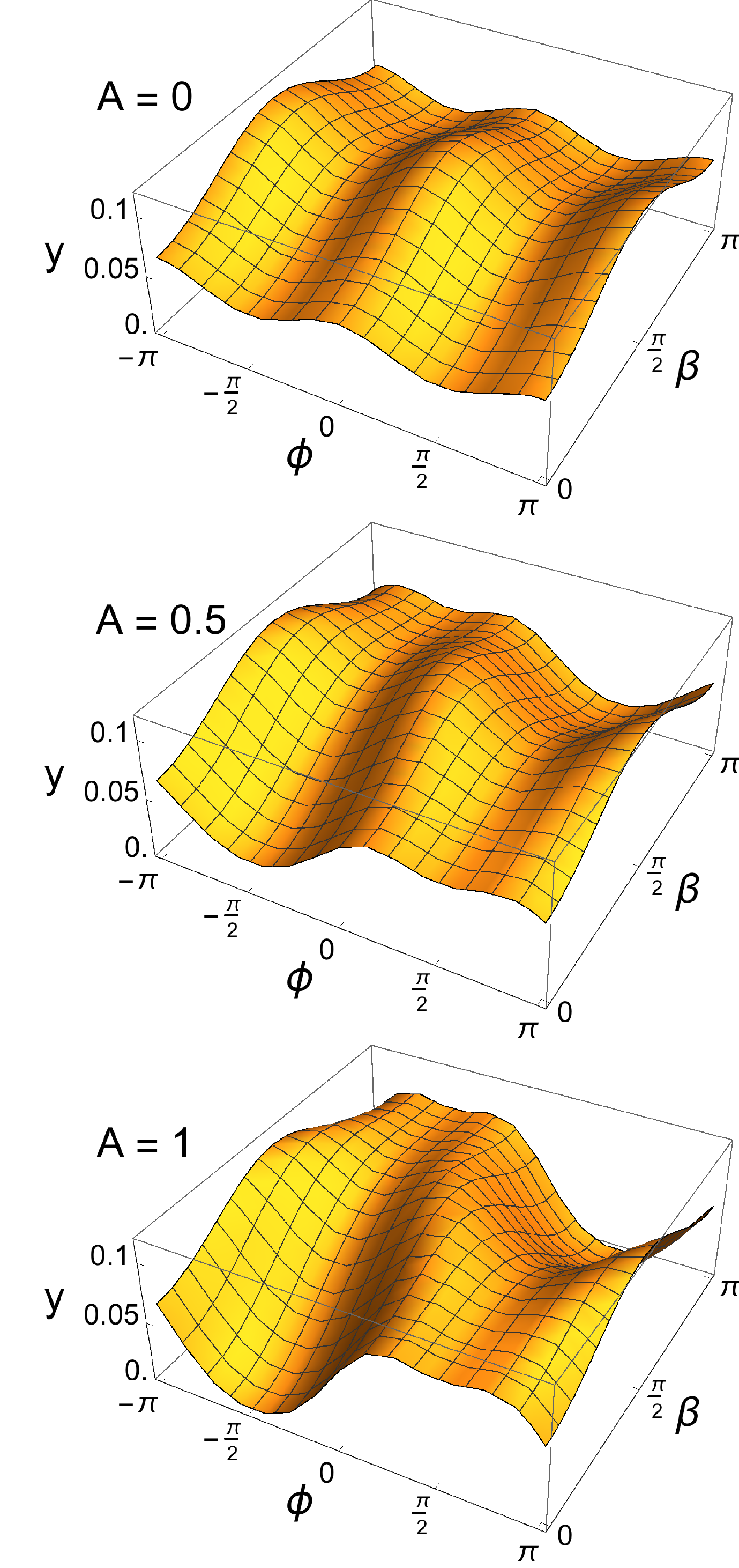}}
\caption{Probability distribution $y(\phi,\beta)$ for $A=0$ (top), $A=0.5$ (middle) and $A=1$ (bottom). For all three cases, $\lambda_\beta$ and $v_2$ are -0.1 and 0.1, respectively. }
  \label{fig:Y3AValues}
\end{figure}        

In Fig.~\ref{fig:Y3AValues}, $y(\phi,\beta)$ is shown for three different $A$ values.  
For a given $\beta$, a finite $A$ value will distort the originally (when $A=0$) symmetric $J/\psi$ production w.r.t. $\phi=0$. 
In principle, the $J/\psi$ yield measured in $(\phi,\beta)$ spaces, if fitted with Eq.(\ref{eq:yProb}), can be used to extract $A$.

Taking the difference in probability between the upper and lower hemispheres,
\bea
\nonumber \Delta Y_{\rm{up}-\rm{down}}^{J/\psi} & = & \int_0^\pi y(\phi) \mathrm{d}\phi - \int_\pi^{2\pi} y(\phi) \mathrm{d}\phi\\
& = & \frac{A}{4\pi}(3-2v_2)(1+\lambda_\beta)\mathrm{cos}\beta .
  \label{eq:Y_ud_a}
\eea
Note that Eq.(\ref{eq:Y_ud_a}) shows that $\Delta Y_{\rm{up}-\rm{down}}^{J/\psi}$ could still be finite even if 
the transverse and longitudinal polarizations have equal possibility ($\lambda_{\beta}=0$). 
Again the difference between the left and right hemispheres (separated by $\phi=\pi/2$) is 
\bea
\nonumber \Delta Y_{\rm{left}-\rm{right}}^{J/\psi} & = &  \int_{\frac{\pi}{2}}^{\frac{3\pi}{2}} y(\phi) \mathrm{d}\phi - \int_{-\frac{\pi}{2}}^{\frac{\pi}{2}} y(\phi) \mathrm{d}\phi\\
& = & 0.
\eea

When the detection efficiency is finite, Eq.(\ref{eq:yProb}) can be simply multiplied by $\epsilon(\beta)\epsilon(\phi)$, 
where $\epsilon(\beta)$ and $\epsilon(\phi)$ are efficiency functions of $\beta$ and $\phi$, respectively.
With a finite efficiency, $\Delta Y_{\rm{up}-\rm{down}}^{J/\psi}$ does not take the simple form of Eq.(\ref{eq:Y_ud_a}) any more.
However, similar to the case of photons, a finite efficiency $\epsilon(\beta)$ will not cause an artificial, finite signal 
in $\Delta Y_{\rm{up}-\rm{down}}^{J/\psi}$. Here we assume that the (in)efficiency as a function of $\phi$ angle can be easily compensated 
in experiments, as long as the RP distribution and the azimuthal angle distribution of particles of interest in the laboratory 
frame are flattened, which is a common practice when analyzing anisotropic flow~\cite{PoskanzerVoloshin}.

\subsection{\label{sec:Jpsi2}Additional asymmetry in yield w.r.t RP between longitudinal and transverse polarizations}

A more general physics case would not exclude the possibility that the portions of 
the longitudinal and transverse polarizations also vary w.r.t. the RP. 
To take this effect into account, one can introduce the RP dependence of $\lambda_{\beta}$ in Sec.~\ref{sec:Jpsi1} as:
\bea
\lambda_{\beta}(\phi) = \lambda_{\beta}^0 (1+2u\, \mathrm{sin}\phi).
\eea
The corresponding $\Delta Y_{\rm{up}-\rm{down}}^{J/\psi}$ becomes
\be
\Delta Y_{\rm{up}-\rm{down}}^{J/\psi} =\\ 
\frac{3-2v_2}{8\pi}[2A(1+\lambda_{\beta}^0)\mathrm{cos}\beta+\lambda_{\beta}^0 u(1+3\mathrm{cos}2\beta)]
\ee
and $\Delta Y_{\rm{left}-\rm{right}}^{J/\psi}$ remains zero owing to the symmetry w.r.t $\phi=\pi/2$ plane.

\vspace{-0.5cm}
\section{Discussion and Summary}
We have shown that for massless photons, as well as vector mesons decaying into two daughters, 
the difference in the circular polarization preference between upper- and lower-hemispheres could be feasibly measured. 
We propose a scheme to carry out the measurement that quantifies the effect. 
In practice, there are other complications that need to be considered in experiments. 
For example, $\pi_0$ decays may dominate the photon production, effectively diluting the signal. 
In the end it is a competition between the strength of the signal and the dilution effect. 
Similar dilution effects can happen to vector mesons due to the background contamination. 
The correction for dilution depends on the detailed experimental setup, which varies from one experiment to another, 
and is beyond the scope of this paper.

\vspace{-0.2cm}
\begin{acknowledgments}
We'd like to thank D. Kharzeev, M. Lisa ,Y. Yin, and Y. Zhang for fruitful discussions. We thank Y. Yin and Y. Zhang for reading the manuscript and providing comments. A.T. was supported by the US Department of Energy under Grants No. DE-AC02-98CH10886 and No. DE-FG02-89ER40531. G.W. was supported by the US Department of Energy under Grants No. DE-FG02-88ER40424.
\end{acknowledgments}

\end{document}